\documentclass{article}
\usepackage{spconf,amsmath,graphicx}
\usepackage{graphicx}
\usepackage{amsmath}
\usepackage{amssymb}
\usepackage{epstopdf}
\usepackage{algorithmic}
\usepackage{subfigure}
\usepackage{pseudocode}
\usepackage{fancybox}
\usepackage{multicol}
\usepackage{verbatim}
\usepackage{caption}
\usepackage{bm}
\usepackage[pdfborder={1 1 1},pdfborderstyle={/S/U/W 1}]{hyperref}

\title{Adaptive structured low rank algorithm for MR image recovery}
\name{Author(s) Name(s)\thanks{Thanks to XYZ agency for funding.}}
\address{Author Affiliation(s)}
\name{Yue Hu$^{\star \thanks{This work is supported by China NSFC 61501146, Natural Science Foundation of Heilongjiang F2016018, NIH 1R01EB019961-01A1 }}$ \qquad Xiaohan Liu$^{\star}$ \qquad Mathews Jacob$^{\dagger}$}
\address{$^{\star}$ Department of Electronics and Information Technology, Harbin Institute of Technology, Harbin, China. \\
  $^{\dagger}$Department of Electrical and Computer Engineering, University of Iowa, IA, USA}
\begin{document}
%
\maketitle
\begin{abstract}

We introduce an adaptive structured low rank algorithm to recover MR images from their undersampled Fourier coefficients. The image is modeled as a combination of a piecewise constant component and a piecewise linear component. The Fourier coefficients of each component satisfy an annihilation relation, which results in a structured Toeplitz matrix. 
We exploit the low rank property of the matrices to formulate a combined regularized
optimization problem, which can be solved efficiently. Numerical experiments indicate that the proposed algorithm provides improved recovery performance over the previously proposed algorithms.
\end{abstract}
\section{Introduction}
\label{sec:intro}
Recovering image data from their noisy partial measurements has been a critical research topic in a wide range of imaging applications including biomedical imaging, remote sensing, and microscopy. The common method is to formulate the image recovery problem as an optimization problem which is the combination of data consistency and regularization term. Conventional regularization penalties include $L_1$ sparsity or smoothness priors . Recently, structured low rank matrix priors have been introduced as powerful alternatives due to their improvement in image reconstruction quality \cite{jin2016general,haldar2014low,ongie2016off}.


Structured low rank matrix algorithms are based on annihilation relationship between the Fourier coefficients of the image and a large set of finite impulse response filter \cite{pan2014sampling,vetterli2002sampling}. 
These algorithms are inspired from the finite-rate-of-innovation (FRI) framework \cite{vetterli2002sampling,maravic2005sampling}. 
However, the direct extension of FRI model to natural image did not work well. Ongie {\em et~al.} overcome the challenges by presenting an improved signal model based on a class of piecewise smooth functions \cite{ongie2016off,ongie2015recovery}. This new model lifted the Fourier samples of signals into structured low rank matrix, and the reconstruction of the signal translates to the problem of matrix completion. The annihilation property results in a convolutional structured low rank matrix, which is built from the Fourier coefficients of the image. Researchers have shown that the structured low rank matrix algorithms can provide improved reconstruction performance than standard total variation methods \cite{ongie2015recovery,ongie2016fast}.


In this paper, we model an MR image as the combination of a piecewise constant component and a piecewise linear component. For the piecewise constant component, 
the Fourier coefficients of the gradient of the component satisfy the annihilation relation. We can thus build a structured Toeplitz matrix, which can be proved to be low rank. Similarly, we can obtain a structured low rank matrix from the Fourier coefficients of the second order partial derivatives of the piecewise linear component. By introducing the adaptive method, both the edges and the smooth regions of the image can be accurately recovered. In order to solve the corresponding optimization problem, we adapt the Generic Iteratively Reweighted Annihilating Filter (GIRAF) algorithm proposed in \cite{ongie2016fast}, which is based on a half-circulant approximation of the Toeplitz matrix. This algorithm alternates between the estimation of the annihilation filter of the image, and the computation of the image anniihilated by the filter in a least squares formulation. We investigate the performance of the algorithm in the context of compressed sensing MR images reconstruction. Experiments show that the proposed method is capable of providing more accurate recovery results than the state of the art algorithms.

\section{Background}
Consider the general model for a 2-D piecewise smooth image $\rho(\mathbf r)$ at the spatial location $\mathbf r=(x,y)\in\mathbb Z^2$:
\begin{equation}
\label{eq:piecesmooth}
\rho(\mathbf r)=\sum_{i=1}^{N}g_i(\mathbf r)\chi_{\Omega_i}(\mathbf r)
\end{equation}
where $\chi_{\Omega_i}$ is a characteristic function of the set $\Omega_i$ and the functions $g_i(\mathbf r)$ are smooth polynomial functions which vanish with a collection of differential operators $\mathbf D=\{D_1,...,D_N\}$ within the region $\Omega_i$. 
We assume that a bandlimited trigonometric polynomial function $\mu(\mathbf r)$ vanishes on the edge set $\partial \Omega=\bigcup _{i=1}^N\partial \Omega_i$ of the image:
\begin{equation}\label{eq:mu}
\mu(\mathbf r)=\sum_{\mathbf k \in \Delta_1}c[\mathbf k]e^{j2\pi\langle\mathbf k,\mathbf r\rangle}
\end{equation}
where $c[\mathbf k]$ denotes the Fourier coefficients of $\mu$ and $\Delta_1$ is any finite set of $\mathbb Z^2$.
According to \cite{ongie2015recovery}, the family of functions in (\ref{eq:piecesmooth}) is a general form including many common image models by choosing different set of differential operators $\mathbf D$.

For example, for a piecewise constant image $\rho_1(\mathbf r)$, the first order partial derivative of the image $\mathbf D_1\rho_1=\bm{\nabla} \rho_1=(\partial_x \rho_1,\partial_y \rho_1)$ is annihilated by multiplication with $\mu$ in the spatial domain, i.e., $\mu\bm{\nabla}\rho_1=0 $. The multiplication in spatial domain translates to the convolution in Fourier domain, by which we can formulate the annihilation property as a matrix multiplication:
:
    \begin{equation}
    {\cal T}_1(\hat \rho_1)\mathbf c=\left[\begin{array}{c}
    {\cal T}_x(\hat\rho_1)\\{\cal T}_y(\hat \rho_1)\end{array}\right]\mathbf c=\mathbf 0
    \end{equation}
    where ${\cal T}_1(\hat\rho_1)$ is a Toeplitz matrix built from the entries of $\hat \rho_1$, the Fourier coefficients of $\rho_1$. ${\cal T}_x(\hat\rho_1)$, ${{\cal T}_y(\hat\rho_1)}$ are matrices derived from $k_x\hat\rho_1[\mathbf k]$ and $k_y\hat\rho_1[\mathbf k]$, omitting the irrelevant factor $j2\pi$. Here $\mathbf c$ is the vectorized version of the filter $c[\mathbf k]$, supported in $\Delta_1$. 
    Consequently, we can obtain:
    \begin{equation}
    \hat\rho_1[\mathbf k]*c_1[\mathbf k]=0, \;\mathbf k\in \Gamma
    \end{equation}
    Here $c_1[\mathbf k]=c[\mathbf k]*h[\mathbf k]$, where $h[\mathbf k]$ is any FIR filter. Note that $\Delta_1$ is smaller than $\Gamma$, the support of $c_1$. Thus, if we take a larger filter size than the minimal filter $c[\mathbf k]$, the annihilation matrix will have a larger null space. Therefore, ${\cal T}_1(\hat\rho_1)$ is a low rank matrix. The method corresponding to this case is referred to as the 1st order structured low rank algorithm for simplicity.

    Similarly, for a piecewise linear image $\rho_2$, the second order partial derivatives of the image satisfy the annihilation property $\mu^2 \mathbf D_2\rho_2=0$, where $\mathbf D_2\rho_2=(\partial^2_{xx}\rho_2,\partial^2_{xy}\rho_2,\partial^2_{yy}\rho_2)$. Thus the annihilation property in this case can be written in the matrix form as:
    \begin{equation}
 {\cal T}_2(\hat\rho_2)\mathbf d=\left[\begin{array}{c}
    {\cal T}_{xx}(\hat\rho_2)\\{\cal T}_{xy}(\hat \rho_2)\\{\cal T}_{yy}(\hat\rho_2)\end{array}\right]\mathbf d=\mathbf 0
 \end{equation}
 where 
 ${\cal T}_{xx}(\hat\rho_2)$, ${{\cal T}_{xy}(\hat\rho_2)}$, and ${{\cal T}_{yy}(\hat\rho_2)}$ are matrices built from $k_x^2\hat\rho_2[\mathbf k]$, $k_xk_y\hat\rho_2[\mathbf k]$, and $k_y^2\hat\rho_2[\mathbf k]$, omitting the insignificant factor; $\mathbf d$ is the vector of $d[\mathbf k]$, the Fourier coefficients of $\mu^2$. Here ${\cal T}_2(\hat\rho_2)$ can also be proved to be a low rank matrix. The method exploiting the low rank property of ${\cal T}_2(\hat\rho_2)$ is referred to as the 2nd order structured low rank method.

\section{Adaptive structured low rank image recovery Algorithm}

We assume that the recovery of MR images from their undersampled measurements can be modeled as:
\begin{equation}
\mathbf b={\cal A}(\hat \rho)+\eta
\end{equation}
where ${\cal A}$ is the measurement operator corresponding to Fourier undersampling of $\hat \rho$, and $\eta$ is the zero mean white Gaussian noise.


We are interested in decomposing an MR image $\rho$ into two components $\rho=\rho_1+\rho_2$, such that $\rho_1$ represents the piecewise constant component of $\rho$, while $\rho_2$ represents the piecewise linear component of $\rho$. We consider the framework of a combined regularization procedure. Specifically, we attempt to solve the following optimization problem:
\begin{align}\label{eq:optprob}
\{\hat\rho_1^\star,\hat\rho_2^\star\}=\arg\min_{\hat\rho_1,\hat\rho_2}
\lambda_1\|{\cal T}_1(\hat\rho_1)\|_p+\lambda_2\|{\cal T}_2(\hat\rho_2)\|_p \nonumber\\
+\|{\cal A}(\hat\rho_1+\hat\rho_2)-\mathbf b \|^2
\end{align}
Here ${\cal T}_i(\hat\rho_i)$ ($i=1,2$) are the structured Toeplitz matrices in the lifted domain. $\|\cdot\|_p$ is the Schatten $p$ norm ($0< p\leq 1$), defined for an arbitrary matrix $\mathbf X$ as $\|\mathbf X\|_p = \frac{1}{p}\mbox{Tr}[(\mathbf X^*\mathbf X)^{\frac{p}{2}}]=\frac{1}{p}\sum\limits_i\sigma_i^p$, where $\sigma_i$ are the singular values of $\mathbf X$. $\lambda_1$ and $\lambda_2$ are regularization parameters which balance the data consistency and the degree to which ${\cal T}_1(\hat\rho_1)$ and ${\cal T}_2(\hat\rho_2)$ are low rank.

\label{sec:pagestyle}

We apply the iterative reweighted least squares (IRLS) algorithm to solve the optimization problem (\ref{eq:optprob}). Based on the equation $\|\mathbf X\|_p=\|\mathbf X\mathbf H^{\frac{1}{2}}\|_F^2$, where $\mathbf H=(\mathbf X^*\mathbf X)^{\frac{p}{2}-1}$, let $\mathbf X={\cal T}_i(\hat\rho_i)$ ($i=1,2$), (\ref{eq:optprob}) becomes:
\begin{align}\label{eq:optprob1}
\{\hat\rho_1^\star,\hat\rho_2^\star\}=&\arg\min_{\hat\rho_1,\hat\rho_2}
\lambda_1\|{\cal T}_1(\hat\rho_1)\mathbf H_1^{\frac{1}{2}}\|^2_F+\lambda_2\|{\cal T}_2(\hat\rho_2)\mathbf H_2^{\frac{1}{2}}\|^2_F\nonumber\\
&+\|{\cal A}(\hat\rho_1+\hat\rho_2)-\mathbf b \|^2
\end{align}
In order to solve (\ref{eq:optprob1}), we can use an alternating minimization scheme, which alternates between the following subproblems: updating the weight matrices $\mathbf H_i$ ($i=1,2$), and solving a weighted least squares problem. Specifically, at $n$th iteration, we compute:
\begin{equation}\label{eq:Hi}
\mathbf H_{i,n}=[{\cal T}_i(\hat\rho_{i,n})^*{\cal T}_i(\hat\rho_{i,n})+\epsilon_n\mathbf I]^{\frac{p}{2}-1}
\end{equation}
\begin{align}\label{eq:leastsquare}
\{\hat\rho_{1,n},\hat\rho_{2,n}\}=&\arg\min_{\hat\rho_1,\hat\rho_2}
\lambda_1\|{\cal T}_1(\hat\rho_1)\mathbf H_{1,n}^{\frac{1}{2}}\|^2_F\nonumber\\
&+\lambda_2\|{\cal T}_2(\hat\rho_2)\mathbf H_{2,n}^{\frac{1}{2}}\|^2_F
+\|{\cal A}(\hat\rho_1+\hat\rho_2)-\mathbf b \|^2
\end{align}
where $\epsilon_n\rightarrow0$ is a small factor used to stabilize the inverse. We now show how to efficiently solve the subproblems.
\subsection{Update of least squares}

First, let ${\mathbf H}_1=[\mathbf h_1^{(1)},...,\mathbf h_1^{(N)}]$, ${\mathbf H}_2=[\mathbf h_2^{(1)},...,\mathbf h_2^{(M)}]$, we rewrite the least squares problem (\ref{eq:leastsquare}) as follows:
\begin{align}\label{eq:leastsquare1}
&\min_{\hat\rho_1,\hat\rho_2}
\lambda_1\sum_{l=1}^N\|{\cal T}_1(\hat\rho_1)\mathbf h_{1}^{(l)}\|^2_F+\lambda_2\sum_{m=1}^M\|{\cal T}_2(\hat\rho_2)\mathbf h_{2}^{(m)}\|^2_F\nonumber\\
&+\|{\cal A}(\hat\rho_1+\hat\rho_2)-\mathbf b \|^2
\end{align}
We now focus on the update of $\hat\rho_1$. The update of  $\hat\rho_2$ can be derived likewise. From the structure property of ${\cal T}_1(\hat\rho_1)$ and the convolution relationship, we can obtain:
\begin{align}\label{eq:1stsimp}
{\cal T}_1(\hat\rho_1)\mathbf h_1^{(l)}&={\cal P}_{\Gamma_1}(\mathbf M_1\hat\rho_1*\mathbf h^{(l)}_1)={\cal P}_{\Gamma_1}(\mathbf h^{(l)}_1*\mathbf M_1\hat\rho_1)\nonumber\\&=\mathbf P \mathbf C_1^{(l)}\mathbf M_1\hat\rho_1,l=1,...,N
\end{align}
where $\mathbf C_1^{(l)}$ denotes the linear convolution by $\mathbf h_1^{(l)}$, ${\cal P}_{\Gamma_1}$ is the projection of the convolution to a finite set $\Gamma_1$ of the valid $k$ space index, which is expressed by the matrix $\mathbf P$. $\mathbf M_1$ is the linear transformation in $k$ space, which is multiplication by the 1st order Fourier derivatives $j2\pi k_x$ and $j2\pi k_y$, referred to as the gradient weight lifting case.
We can approximate $\mathbf C_1^{(l)}$ by a circular convolution by $\mathbf h_1^{(l)}$ on a sufficiently large convolution grid. Then, we can obtain $\mathbf C_1^{(l)}=\mathbf F\mathbf S_1^{(l)}\mathbf F^*$, where $\mathbf F$ is the 2-D DFT and $\mathbf S_1^{(l)}$ is a diagonal matrix representing multiplication by the inverse DFT of $\mathbf h_1^{(l)}$.
Assuming $\mathbf P^*\mathbf P\approx \mathbf I$, we can thus rewrite the first term in (\ref{eq:leastsquare1}) as:
\begin{align}
\lambda_1\sum_{l=1}^N\|\mathbf P\mathbf C_1^{(l)}\mathbf M_1\hat\rho_1\|^2&=\lambda_1\hat\rho_1^*\mathbf M_1^*\mathbf F\underbrace{\sum_{l=1}^N\mathbf S_1^{(l)*}\mathbf S_1^{(l)}}_{\mathbf S_1}\mathbf F^*\mathbf M_1\hat\rho_1\nonumber\\
&=\lambda_1\|\mathbf S_1^{\frac{1}{2}}\mathbf F^*\mathbf M_1\hat\rho_1\|^2
\end{align}
where $\mathbf S_1$ is a diagonal matrix with entries $\sum_{l=1}^N|\mu_l(\mathbf r)|^2$, where $\mu_l(\mathbf r)$ is the trigonometric polynomial of inverse Fourier transform of $\mathbf h_1^{(l)}$.

Similarly, the second term in (\ref{eq:leastsquare1}) can be rewritten as $\lambda_2\|\mathbf S_2^{\frac{1}{2}}\mathbf F^*\mathbf M_2\hat\rho_2\|^2$. Therefore, we can reformulate the optimization problem (\ref{eq:leastsquare1}) as:
\begin{align}\label{eq:newprob}
\min_{\hat\rho_1,\hat\rho_2}
\lambda_1\|\mathbf S_1^{\frac{1}{2}}\mathbf y_1\|^2_F+&\lambda_2\|\mathbf S_2^{\frac{1}{2}}\mathbf y_2\|^2_F
+\|{\cal A}(\hat\rho_1+\hat\rho_2)-\mathbf b \|^2 \nonumber\\
&\mbox{s.t.}\;\mathbf F\mathbf y_1=\mathbf M_1\hat\rho_1,\;\mathbf F\mathbf y_2=\mathbf M_2\hat\rho_2
\end{align}

The above constrained problem can be efficiently solved using the alternating directions method of multipliers (ADMM) algorithm \cite{esser2009applications}, which yields to solving the following subproblems:
\begin{equation}\label{y1}
\mathbf y_1^{(n)}=\min_{\mathbf y_1}\|\mathbf S_1^\frac{1}{2}\mathbf y_1\Vert_2^2+\gamma_1\|\mathbf q_1^{(n-1)}+\mathbf F^*\mathbf M_1{\hat\rho_1}^{(n-1)}-{\mathbf y}_1\Vert_2^2
\end{equation}

\begin{equation}\label{y2}
\mathbf y_2^{(n)}=\min_{\mathbf y_2}\|\mathbf S_2^\frac{1}{2}\mathbf y_2\Vert_2^2+\gamma_2\|\mathbf q_2^{(n-1)}+\mathbf F^*\mathbf M_2{\hat\rho_2}^{(n-1)}-\mathbf y_2\Vert_2^2
\end{equation}

\begin{equation}\label{rho1}
\hat\rho_1^{(n)}=\min_{\hat\rho_1}\|{\cal A}(\hat\rho_1+\hat\rho_2)-\mathbf b\Vert_2^2+\gamma_1\lambda_1\|\mathbf q_1^{(n-1)}+\mathbf F^*\mathbf M_1\hat\rho_1-\mathbf y_1^{(n)}\Vert_2^2\
\end{equation}

\begin{equation}\label{rho2}
\hat\rho_2^{(n)}=\min_{\hat\rho_2}\|{\cal A}(\hat\rho_1+\hat\rho_2)-\mathbf b\Vert_2^2
+\gamma_2\lambda_2\|\mathbf q_2^{(n-1)}+\mathbf F^*\mathbf M_2\hat\rho_2-\mathbf y_2^{(n)}\Vert_2^2
\end{equation}

\begin{equation}
\mathbf q_i^{(n)}=\mathbf q_i^{(n-1)}+\mathbf F^*\mathbf M_{i}\hat\rho_{i}^{(n-1)}-\mathbf y_{i}^{(n)}\quad i=1,2
\end{equation}
where $\mathbf q_{i}$ ($i=1,2$) represent the vectors of Lagrange multipliers, and $\gamma_{i}$ ($i=1,2$)  are fixed parameters tuned to improve the conditioning of the subproblems. Subproblems (\ref{y1}) to (\ref{rho2}) are quadratic and thus can be solved easily as follows:
\begin{equation}
\mathbf y_1^{(n)}=(\mathbf S_1+\gamma_1\mathbf I)^{-1}[\gamma_1(\mathbf q_1^{(n-1)}+\mathbf F^*\mathbf M_1\hat\rho_1^{(n-1)})]
\end{equation}
\begin{equation}
\mathbf y_2^{(n)}=(\mathbf S_2+\gamma_2\mathbf I)^{-1}[\gamma_2(\mathbf q_2^{(n-1)}+\mathbf F^*\mathbf M_2 \hat\rho_2^{(n-1)})]
\end{equation}
\begin{equation}
\begin{split}
\hat\rho_1^{(n)}=({\cal A}^*{\cal A}+\gamma_1\lambda_1\mathbf M_1^*\mathbf M_1)^{-1}[&\gamma_1\lambda_1(\mathbf M_1^*\mathbf F)(\mathbf y_1^{(n)}-\mathbf q_1^{(n-1)})\\
&+{\cal A}^*\mathbf b-{\cal A}^*{\cal A}\hat\rho_2^{(n-1)}]
\end{split}
\end{equation}
\begin{equation}
\begin{split}
\hat\rho_2^{(n)}=({\cal A}^*{\cal A}+\gamma_2\lambda_2\mathbf M_2^*\mathbf M_2)^{-1}[&\gamma_2\lambda_2(\mathbf M_2^*\mathbf F)(\mathbf y_2^{(n)}-\mathbf q_2^{(n-1)})\\
&+{\cal A}^*\mathbf b-{\cal A}^*{\cal A}\hat\rho_1^{(n-1)}]
\end{split}
\end{equation}

\subsection{Update of weight matrices}

We now show how to update the weight matrices in (\ref{eq:Hi}) 
efficiently based on the GIRAF method \cite{ongie2016off}. Let $(\mathbf V_i,\bm{\Lambda}_i)$ denote the eigen-decomposition of ${\cal T}_i(\hat\rho_i)^*{\cal T}_i(\hat\rho_i)$ ($i=1,2$), where $\mathbf V_i$ is the orthogonal basis of eigenvectors $\mathbf v_i$, and $\bm{\Lambda_i}$ is the diagonal matrix of eigenvalues $\lambda_k$ satisfying ${\cal T}_i(\hat\rho_i)=\mathbf V_i \bm{\Lambda}_i \mathbf V_i^*$. Then we can rewrite the weight matrices $\mathbf H_1$ and $\mathbf H_2$ as:
\begin{equation}
\mathbf H_i=[\mathbf V_i(\bm{\Lambda}_i+\epsilon\mathbf I)\mathbf V_i^*]^{\frac{p}{2}-1}=\mathbf V_i(\bm{\Lambda}_i+\epsilon\mathbf I)^{\frac{p}{2}-1}\mathbf V_i^*,\;i=1,2
\end{equation}
Thus, one choice of the matrix square root $\mathbf H_i^{\frac{1}{2}}$ is $(\bm\Lambda_i+\epsilon\mathbf I)^{\frac{p}{4}-\frac{1}{2}}\mathbf V_i^*$.

\section{Results}
\label{sec:typestyle}

The performance of the proposed method is investigated in the context of compressed sensing MR images reconstruction. In the experiments, we assume that the measurements are acquired using variable density random retrospective sampling pattern under different acceleration factors. For each ${\cal A}$ operator, we determine the regularization parameters to obtain the optimized signal-to-noise ratio (SNR) to ensure fair comparisons between different methods. 

\begin{figure}[ht!]
\vspace{-0.6em} \centering
{\hspace{-0.3em}\includegraphics[width= 8.05cm]{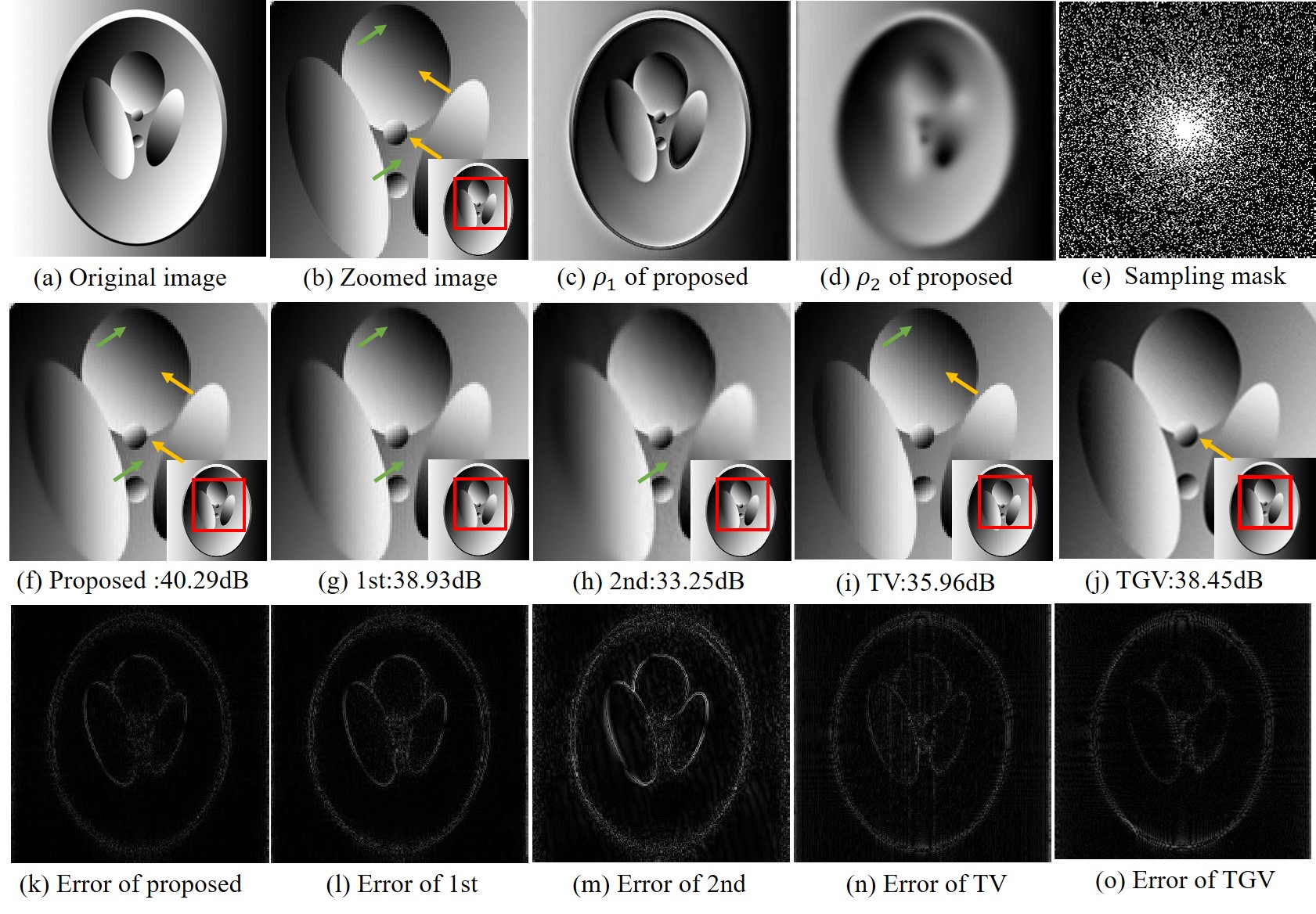}}
\setlength{\fboxrule}{0.5pt}
\caption{Recovery of a piecewise smooth phantom image from the 4-fold undersampled measurements using $15\times 15$ filter size. (a)-(b): The actual and the zoomed version of the image. (c)-(d): The decomposition results. (e): The sampling mask. (f)-(j): Reconstructions using the proposed method, the 1st and 2nd order method, TV, and TGV, respectively. (k)-(o): Error images.}\label{sequentialfig}\vspace{-0.8em} \label{phantom}
\end{figure}

We first study the performance of the proposed method 
for the recovery of a piecewise smooth phantom image from its noiseless k-space data at 4-fold undersampling in Fig. \ref{phantom}. Note that the decomposition results by the propose method indicated in (c) and (d) clearly show the piecewise constant component $\rho_1$ and the piecewise linear component $\rho_2$ of the image. We observe that the proposed method provides lower errors compared with the standard TV, the 1st and 2nd order structured low rank methods, and the total generalized variation (TGV) method \cite{knoll2011second}. The computational time with the GPU implementation for the 1st order method, the proposed method, and TGV were 9.4 sec, 18.3 sec, and 5.2 sec, respectively.


\begin{figure}[ht!]
\vspace{-0.6em} \centering

{\hspace{-0.3em}\includegraphics[width= 7.35cm]{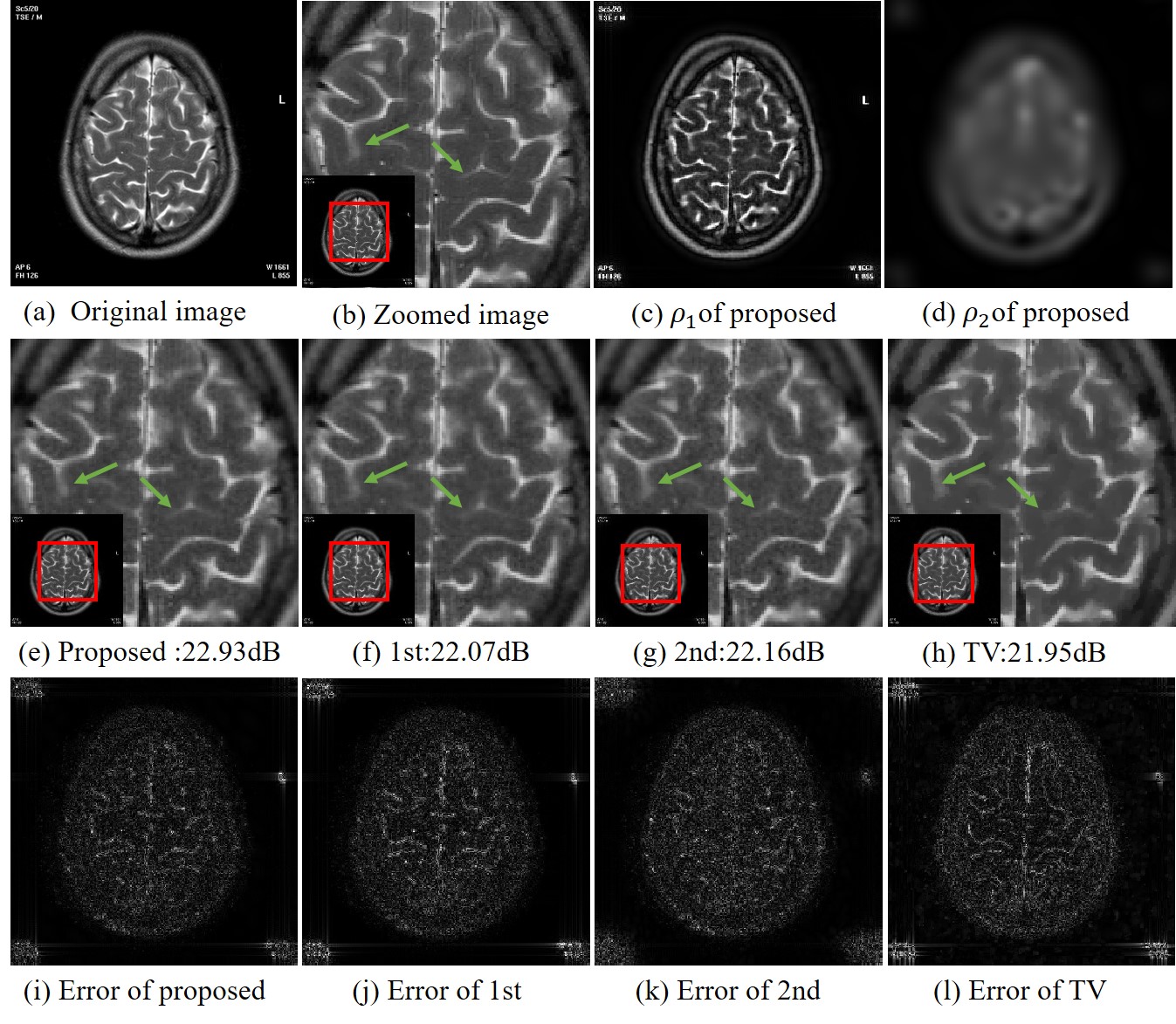}}
\setlength{\fboxrule}{0.5pt}
\caption{Recovery of the brain MR dataset from 5-fold undersampled measurements using $15\times 15$ filter size. (a)-(b): The actual and the zoomed version of the original image. (c)-(d): The decomposition results. (e)-(h): Reconstructions using the proposed method, the 1st and 2nd order method, and standard TV, respectively. (i)-(l): Error images.}\label{sequentialfig}\vspace{-0.8em} \label{brain1}
\end{figure}

\begin{figure}[ht!]
\vspace{-0.2em} \centering

{\hspace{-0.3em}\includegraphics[width= 8.05cm]{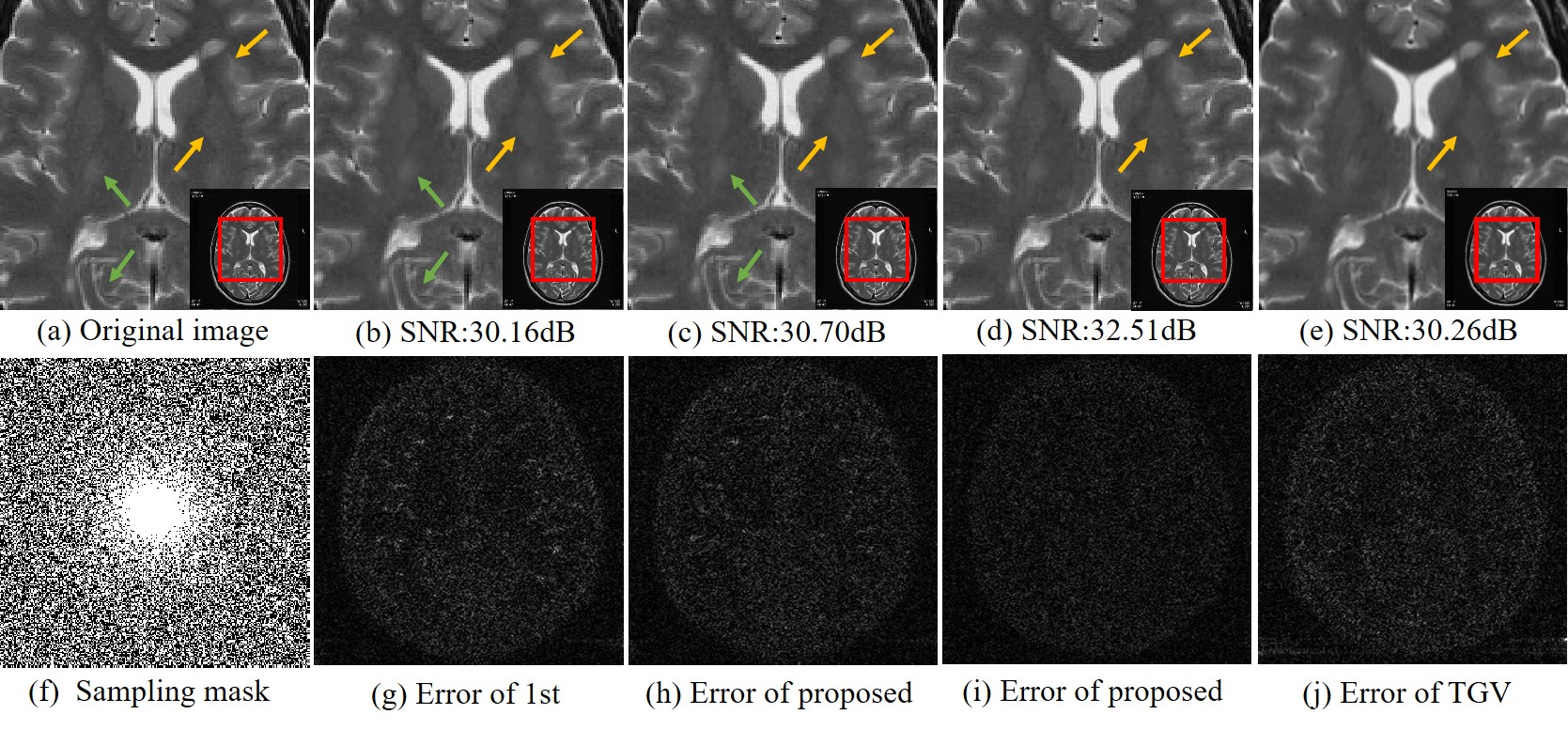}}
\setlength{\fboxrule}{0.5pt}
\caption{Recovery of the brain MR dataset from 2-fold undersamppled measurements. (a): The actual zoomed image. (b): The recovery image using the 1st order scheme with filter size of $15\times 15$. (c)-(d): The reconstructions using the proposed method with filter size of $15\times 15$ and $31\times 31$, respectively. (e): The reconstruction of TGV. (f): The undersampling pattern. (g)-(j): Error images. }\label{sequentialfig}\vspace{-0.8em} \label{brain2}
\end{figure}

In Fig. \ref{brain1}, we demonstrate the performance of the proposed approach 
on the reconstruction of a brain MR image from 5-fold undersampling. The results show that TV leads to patchy recovered image. 
While the proposed method outperforms the other schemes in 
providing more accurate recovered image. In Fig. \ref{brain2}, we demonstrate the effect of the proposed scheme using different filter sizes on the recovery of another brain MR image data at acceleration factor of 2. We observe that as the filter size increases, the proposed method provides better result, indicating the benefits of using larger filters.

\section{Conclusion}
\label{sec:majhead}

We proposed a novel adaptive structured low rank algorithm to recover MR images from their undersampled $k$ space measurements, by the assumption that an MR image can be modeled as the combination of a piecewise constant component and a piecewise linear component. Experiments show that the proposed algorithm provides more accurate recovery results compared with the state of the art approaches.




\bibliographystyle{IEEEbib}
{\footnotesize\bibliography{strings,acl}}

\begin{thebibliography}{10}

\bibitem{jin2016general}
K.~H. Jin, D.~Lee, and J.~C. Ye,
\newblock ``A general framework for compressed sensing and parallel {MRI} using
  annihilating filter based low-rank {H}ankel matrix,''
\newblock {\em IEEE Transactions on Computational Imaging}, vol. 2, no. 4, pp.
  480--495, 2016.

\bibitem{haldar2014low}
J.~P. Haldar,
\newblock ``Low-rank modeling of local $ k $-space neighborhoods ({LORAKS}) for
  constrained {MRI},''
\newblock {\em IEEE Transactions on Medical Imaging}, vol. 33, no. 3, pp.
  668--681, 2014.

\bibitem{ongie2016off}
G.~Ongie and M.~Jacob,
\newblock ``Off-the-grid recovery of piecewise constant images from few
  {F}ourier samples,''
\newblock {\em SIAM Journal on Imaging Sciences}, vol. 9, no. 3, pp.
  1004--1041, 2016.

\bibitem{pan2014sampling}
H.~Pan, T.~Blu, and P.~L. Dragotti,
\newblock ``Sampling curves with finite rate of innovation,''
\newblock {\em IEEE Transactions on Signal Processing}, vol. 62, no. 2, pp.
  458--471, 2014.

\bibitem{vetterli2002sampling}
M.~Vetterli, P.~Marziliano, and T.~Blu,
\newblock ``Sampling signals with finite rate of innovation,''
\newblock {\em IEEE Transactions on Signal Processing}, vol. 50, no. 6, pp.
  1417--1428, 2002.

\bibitem{maravic2005sampling}
I.~Maravic and M.~Vetterli,
\newblock ``Sampling and reconstruction of signals with finite rate of
  innovation in the presence of noise,''
\newblock {\em IEEE Transactions on Signal Processing}, vol. 53, no. 8, pp.
  2788--2805, 2005.

\bibitem{ongie2015recovery}
G.~Ongie and M.~Jacob,
\newblock ``Recovery of piecewise smooth images from few {F}ourier samples,''
\newblock in {\em Sampling Theory and Applications (SampTA), 2015 International
  Conference on}. IEEE, 2015, pp. 543--547.

\bibitem{ongie2016fast}
G.~Ongie and M.~Jacob,
\newblock ``A fast algorithm for structured low-rank matrix recovery with
  applications to undersampled {MRI} reconstruction,''
\newblock in {\em Biomedical Imaging (ISBI), 2016 IEEE 13th International
  Symposium on}. IEEE, 2016, pp. 522--525.

\bibitem{esser2009applications}
E.~Esser,
\newblock ``Applications of {L}agrangian-based alternating direction methods
  and connections to split {B}regman,''
\newblock {\em CAM report}, vol. 9, pp. 31, 2009.

\bibitem{knoll2011second}
Florian Knoll, Kristian Bredies, Thomas Pock, and Rudolf Stollberger,
\newblock ``Second order total generalized variation ({TGV}) for {MRI},''
\newblock {\em Magnetic resonance in medicine}, vol. 65, no. 2, pp. 480--491,
  2011.

\end{thebibliography}

\end{document}